\input harvmac
\def\half{{1 \over 2}}

\def\>{{\rangle}}
\def\<{{\langle}}

\def\p{{\partial}}

\def\l{{\lambda}}

\def\a {{\alpha}}
\def\b {{\beta}}

\def\g {{\gamma}}
\def\d {{\delta}}

\def\we{{\wedge}}
\def\e {{\epsilon}}

\def \t {{\theta}}

\Title{
\vbox{
\hbox{IFT-P.072/97}  
\hbox {QMW-97-36} \hbox{LPTENS 97/58}  \hbox{hep-th/9712007}
} }
{\vbox{\centerline{\bf 
Manifestly Covariant Actions for}
\bigskip\centerline{\bf 
D=4 Self-Dual Yang-Mills and D=10
Super-Yang-Mills}}}
\bigskip\centerline{Nathan Berkovits}
\bigskip\centerline{Instituto
de F\'{\i}sica Te\'orica, Univ. Estadual Paulista}
\centerline{Rua Pamplona 145, S\~ao Paulo, SP 01405-900, BRASIL}
\centerline{e-mail: nberkovi@ift.unesp.br}
\vskip .1in
\bigskip\centerline{Chris Hull}
\bigskip\centerline{Physics Department, Queen Mary and Westfield College,}
\centerline{ Mile End Road, London E1 4NS, U.K.}
\centerline {and}
\centerline{Laboratoire de Physique Th\' eorique, Ecole Normale Sup\' erieure, }
\centerline{24 Rue Lhomond, 75231 Paris Cedex 05, France}
\bigskip\centerline{e-mail: C.M.Hull@qmw.ac.uk}
\vskip .1in

Using an infinite number of fields, we construct actions for
$D=4$ self-dual Yang-Mills with manifest Lorentz invariance and for
$D=10$ super-Yang-Mills with manifest super-Poincar\'e invariance.
These actions are generalizations of the covariant action for the
$D=2$ chiral boson which was first studied by McClain, Wu, Yu and
Wotzasek.  

\Date{November 1997} 

\newsec {Introduction}

When a physical system has symmetries, it is useful for these symmetries
to be manifest in the action. However, there is a prominent example
where this is not straightforward: Lorentz invariance in self-dual systems.
There are currently at least three methods available for 
attacking this problem.

The first method was proposed by Siegel\ref\SH{W.Siegel,
Nucl. Phys. B238 (1984) 307.}\ and introduces
a Lagrange multiplier for the square of the self-duality constraint.
This method has been most successful in the treatment of
chiral scalars, for which all anomalies can be 
cancelled by introducing an auxiliary non-dynamical sector \ref\ch{C.M. Hull,
 Phys. Lett. {\bf 206} (1988) 234.}. 
The resulting theory is a conformal field theory coupled to
{\it two} world-sheet metrics, 
with a conformal invariance associated with each, and has the correct
spectrum. However, 
the partition 
function depends on the moduli for {\it both} metrics and must have
modular invariance for both 
the moduli of the original metric and for the Siegel multiplier field
\ref\chu{C.M. Hull, unpublished.}. Nevertheless, 
a consistent quantum theory emerges, and at least
in some cases, integrating first over the extra moduli gives the desired result.

The second method (MWYW) was developed by McClain, Wu, Yu
\ref\MWY{B. McClain, Y.S. Wu, and
F. Yu, Nucl. Phys. B343 (1990) 689.}
and 
by Wotzasek  \ref\Wot{C. Wotzasek, Phys. Rev. Lett. 66 (1991) 129.}
for the $D=2$ chiral boson. The chirality condition is
a second-class constraint, and after introducing new fields, can
be transformed into a first-class constraint. However, these new
fields satisfy second-class constraints, and one has to continue the
procedure ad infinitum. The final action therefore contains an infinite number
of fields.

This MWYW method was later used to construct $D=4$ Maxwell 
 \ref\MR{I. Martin and A. Restuccia, Phys. Lett. B323 (1994) 311.}
and super-Maxwell \ref\smax{
N. Berkovits, Phys. Lett. B398 (1997) 79.}
actions with manifest electro-magnetic duality, and to construct
covariant actions in $D=4p+2$ for  $2p$-form gauge fields with self-dual
field strengths
 \ref\DH{F.P. Devecchi and M. Henneaux, Phys. Rev. D54 (1996) 1606,
hep-th 960303.} \ref\BK{
I. Bengtsson and A. Kleppe, ``On chiral $p$-forms'', hep-th 9609102.}. 
There is such
a self-dual 4-form gauge field 
in the spectrum of the Type IIB superstring, and
it was shown in  \ref\sft{N. Berkovits, Phys. Lett. B388 (1996) 743.}\
that superstring field theory utilizes the MWYW
method for constructing its action. 

The third method (PST) for constructing covariant actions for self-dual
systems was developed by Pasti, Sorokin and Tonin \ref\psto
{P. Pasti, D. Sorokin and M. Tonin, Phys. Rev. D55 (1997) 6292.} \ref\pstt{
P. Pasti, D. Sorokin and M. Tonin, Phys. Rev. D52 (1995) 4277.}. 
Starting from
an action without manifest covariance, the PST method introduces a
new harmonic-like field which allows the action to be written in
a manifestly covariant form. This method has the advantage
over the MWYW method of using a finite number of fields, but
it has the disadvantage that the resulting action 
involves inverse powers of the harmonic-like field. 

In this paper, we shall use the MWYW method
to construct manifestly covariant actions  
for $D=4$ self-dual Yang-Mills 
and $D=10$ super-Yang-Mills
with arbitrary gauge group.
At the present time, we do not know how to 
construct analogous actions using the PST method. Note that
it was shown in \psto how to obtain the MWYW action from the PST action,
but it does not seem straightforward to obtain the PST action from the
MWYW action. 

In the second section of this paper, we shall begin
by reviewing the MWYW method
for the $D=2$ chiral boson. 
As pointed out by
Nekrassov 
\ref\neka{N. Nekrassov, private commmunication.}, 
this $D=2$ action involving an infinite number of fields
can be understood as a discretized version of the $D=3$
Chern-Simons action for an abelian one-form. 
We will explicitly construct this $D=3$ Chern-Simons
action and show that it is at level $\half$ for a chiral boson at
free-fermion radius. It is interesting to note that 
a level $\half$ Chern-Simons action was recently used
in \ref\wit {E. Witten, ``Five-Brane Effective Action in
M-Theory'', hep-th 9610234.}
for defining the partition function for
a chiral scalar. 

In the third section, we use the MWYW methed to construct 
a $D=4$ self-dual Yang-Mills action with manifest Lorentz-invariance, and
in the fourth section, we use the MWYW method to constuct 
a $D=10$ super-Yang-Mills action with manifest 
super-Poincar\'e invariance. Manifestly covariant actions 
for these two systems
have never previously been constructed. 
Finally, in the fifth section, we make some comments on actions with an 
infinite number of fields. 

\newsec{MWYW Actions for D=2 Chiral Boson }

\subsec{Hamiltonian formalism}

In their original papers, McClain, Wu, Yu \MWY and Wotzasek \Wot used the
Hamiltonian formalism to describe the $D=2$ chiral boson. Although this
formalism is not manifestly Lorentz-covariant, it is useful for the
analysis of physical states. 

For a $D=2$ non-chiral boson described by canonical variables $\phi(x)$
and its momentum $\Pi(x)$, the free action is
${\cal S}= -{\cal H} +\int d^2 x \Pi\p_0\phi$ where 
${\cal H}=\half\int d^2 x (\Pi^2 + (\p_1\phi)^2)$.
The chirality condition 
$$\Pi-\p_1\phi =0$$ 
is a second-class constraint which makes it difficult to implement using
an action principle.  
MWYW convert it to the first-class constraint
$$\Pi -\p_1\phi - \Pi_{(1)} -\p_1\phi_{(1)} =0$$
where $\phi_{(1)}(x)$ and $\Pi_{(1)}(x)$ are canonical
variables for a second scalar boson. 
In order to describe a single chiral boson, one needs to impose in addition
$\Pi_{(1)}+\p_1\phi_{(1)}=0$. This  in turn is a second-class constraint,
which can itself be converted to a first class constraint by
  repeating the procedure and introducing yet another scalar
boson described by the 
canonical variables $\phi_{(2)}(x)$ and $\Pi_{(2)}(x)$. This continues
ad infinitum to produce the action
\eqn\above{
{\cal S}= -{\cal H} +\sum_{n=0}^\infty\int d^2 x \Pi_{(n)} \p_0 \phi_{(n)}}
where
$${\cal H}= \sum_{n=0}^\infty \int d^2 x [
\half (-1)^n (\Pi_{(n)}^2 +(\p_1\phi_{(n)})^2 ) +\l_{(n)} T_{(n)}],$$
$$T_{(n)}= \Pi_{(n)} -\p_1\phi_{(n)} -\Pi_{(n+1)} -\p_1\phi_{(n+1)},$$
$\l_{(n)}$ are the Lagrange multipliers for the first-class constraints
$T_{(n)}=0$, and $\lbrace \phi_{(0)},\Pi_{(0)}\rbrace$ 
are the canonical variables for
the original scalar.

To prove that this action describes a single chiral scalar, one uses
the fact that 
$$\left[ ~T_{(m)} (x_0, x_1)~,~ 
\phi_{(n+1)}(x_0, x'_1)
+ \int^{x'_1} dy ~\Pi_{(n+1)}(x_0, y) ~
 \right]= - \d_{mn} \d (x_1 -x'_1)$$
to gauge $\phi_{(n)}(x_0, x_1)=
- \int^{x_1} dy \Pi_{(n)}(x_0, y) $
for all $n>0$. Together with the $T_{(n)}$ constraints,
this implies the desired conditions:
$$\Pi_{(0)} = \p_1 \phi_{(0)},\quad
\phi_{(n)} =  \Pi_{(n)} =0 ~~for~ n>0.$$
Although one might worry about determining the physical spectrum
when there are an infinite number of fields and gauge-invariances, 
reference \MWY\ confirms this result by performing a careful analysis  
of the BRST cohomology using the OSp(1,1) method. 

\subsec{Lagrangian formalism}

As shown in \BK, the manifestly covariant form of \above\ is
obtained by solving the equations of motion for $\Pi_{(n)}$ which
produces:
\eqn\cov{{\cal S}={1\over{4\pi}}\sum_{n=0}^\infty
\int d^2 x [\half (-1)^n \p_+ \phi_{(n)}\p_- \phi_{(n)} 
+
A_{(2n)}^- (\p_-\phi_{(2n)} +\p_- \phi_{(2n+1)})}
$$ -
A_{(2n+1)}^+ (\p_+\phi_{(2n+1)} +\p_+ \phi_{(2n+2)}) 
+
(A_{(2n)}^- -A_{(2n+2)}^-)A_{(2n+1)}^+ ],$$
where 
$A_{(2n)}^-$ and 
$A_{(2n+1)}^+$ are Lagrange multipliers
which transform like the $0-1$ and $0+1$ components 
of an SO(1,1) vector under Lorentz transformations, 
$\p_\pm =\p_0 \pm \p_1$, 
and our overall normalization for the action is
chosen to reproduce the free fermion radius if $\phi$ is identified
with $\phi +2\pi $. 

This action is invariant under the gauge transformation:
\eqn\gauge
{\delta\phi_{(n)}=\l_{(n-1)}-\l_{(n)},
\quad\delta A^\pm_{(n)}= \p_\mp \l_{(n)},} 
(with $\l_{(-1)} \equiv 0$)
which allows $\phi_{(n)}$ to be algebraically gauged away for $n\geq 1$. 
In this gauge, \cov\ simplifies to 
\eqn\simp{{\cal S}={1\over 4\pi}
\int d^2 x [\half\p_+ \phi_{(0)}\p_-\phi_{(0)}
+  A_{(0)}^- \p_- \phi_{(0)} +\sum_{n=0}^\infty 
( A_{(2n)}^- - A_{(2n+2)}^-) A_{(2n+1)}^+ ].}

The equations of motion for \simp\ are easily found to be
\eqn\sol{\p_-\phi_{(0)} + A_{(1)}^+ = 0,\quad 
\p_-\p_+\phi_{(0)} = \p_-  A^-_{(0)},}
$$ A_{(1)}^+ = A_{(3)}^+ = 
A_{(5)}^+ = ... ,
\quad A_{(0)}^- = A_{(2)}^- =
 A_{(4)}^- = ... .$$ 
If solutions to \sol\  
are required to contain a finite number of non-vanishing
fields, the solutions must satisfy
\eqn\satisfy{
\p_-\phi_{(0)}=0,\quad 0= A_{(1)}^+ =  A_{(3)}^+ =  ... ,\quad
0= A_{(0)}^- = A_{(2)}^- = ...,}
which are the desired conditions.

Although the action of \simp\ might appear too trivial to have any
useful applications, it is interesting to note that superstring
field theory uses this type of action to describe Ramond-Ramond
fields and their coupling to $D$-branes \sft.

\subsec{Discretized Chern-Simons}

The action of \cov\ can be understood geometrically 
\neka
\ref\nek{
A. Gerasimov, N. Nekrassov and S. Shatishvili, work in progress\semi 
A. Alekseev and S. Shatishvili, Comm. Math. Phys. 128 (1990) 197.}
as a discretized version of the 
abelian Chern-Simons action in three dimensions.
This is easily seen by writing \cov\ as 
$${\cal S}={1\over {4\pi}}\sum_{n=0}^\infty
\int d^2 x [ \half (-1)^n \p_+ A^2_{(n)}\p_- A^2_{(n)} (\delta x_2)^2
+
(A_{(2n)}^- -A_{(2n+2)}^- )A_{(2n+1)}^+ $$
\eqn\cs{ 
+ A_{(2n)}^- (\p^+ A^2_{(2n)} +\p^+ A^2_{(2n+1)}) \delta x_2 -
A_{(2n+1)}^+ (\p^- A^2_{(2n)} +\p^- A^2_{(2n+1)}) \delta x_2 ],}
where $A_{(n)}^2=\phi_{(n)}/\delta x_2$ and $\delta x_2$ is a small
positive constant. Now introduce a third coordinate $x_2$ which is discretized
to take 
non-negative values $n\delta x_2$ and define 
$A^\mu (x_0,x_1,n \delta x_2)= A^\mu_{(n)} (x_0,x_1)$ for $\mu=0,1,2$.

In the continuum limit 
as $\delta x_2 \to 0$, the first term of \cs\
drops out and 
the limit of the  other terms become the  abelian Chern-Simons
action 
\eqn\csa{{\cal S} = {1\over{4\pi}} \epsilon^{\mu\nu\rho}
\int_0^\infty dx_2 \int d^2 x   A_\mu \p_\nu A_\rho
= {1\over{4\pi}} \epsilon^{\mu\nu\rho}
\int d^3 x  A_\mu \p_\nu A_\rho}
where the three-dimensional integration region 
is over the half-volume $x_2 \geq 0$. 
Note that the normalization factor of ${1\over {4\pi}}$ implies that the
Chern-Simons theory is at level $\half$ \wit. (The normalization is fixed
by requiring that $\phi$ is at the free fermion radius and that 
the gauge transformations of \gauge\ are the discretized version of
$\d A_\mu =\p_\mu \lambda$.) 

In this discretized
$D=3$ form of the action, it is easy to explain the condition that
solutions should contain only a finite number of non-vanishing
fields. It is just the usual asymptotic condition that the fields 
should vanish at infinity.

The identification of \cov\ and \csa\ presents a puzzle since
the continuum
version of the Chern-Simons
theory is not chiral, whereas \satisfy\ is.
The chirality is hidden in the passage from discrete to continuous
fields near the boundary. In the discrete version, $A^+$ is defined
at $x_2 = (2n+1) \d x_2$, but not at $x_2=2n \d x_2$. To define $A^+$ at
$x_2 =2n\d x_2$, one should take the average value of $A^+_{(2n-1)}$ and 
$A^+_{(2n+1)}$. Since $A^+_{(-1)}$ is undefined, this means 
fixing $A^+(x_0, x_1, 0)$ = $A^+_{(1)} (x_0, x_1)$. So
in the continuum version, 
$$0= A^+(x_0,x_1, \d x_2) - A^+(x_0, x_1, 0)
= \d x_2 \p_2 A^+(x_0, x_1, 0).$$
But $\p _2 A^+ = \p^+ A_2 $ on-shell, so
$$0 = \d x_2 \p^+ A_2 (x_0, x_1, 0) = \p^+ \phi_{(0)},$$
which is the desired chirality 
condition, arising from a chiral boundary condition. 

The $D=3$ version of the MWYW action for a chiral scalar is easily generalized
to a $D=4p+3$ version of the MWYW action for a chiral $2p$-form. 
In this case, $\phi$ is a $2p$-form, $A$ is a $2p+1$-form defined such that 
$A_{\mu_1 ... \mu_{2p+1}}$ are the Lagrange multipliers for $\mu_i =0$ to
$4p+1$, 
$$A_{\mu_1 ... \mu_{2p} {4p+2}} = \phi_{\mu_1 ... \mu_{2p}} 
/\d x_{4p+2 }, $$
and the appropriate action is a discretized version of the $D=4p+3$
Chern-Simons action:
\eqn\csg{
{\cal S} = 
{1\over{4\pi}} \epsilon^{\mu_1 ... \mu_{2p+1} \nu \rho_1 ...\rho_{2p+1}}
\int_0^\infty dx_{4p+2} \int d^{4p+2} x A_{\mu_1 ... \mu_{2p+1}} 
\p_\nu A_{\rho_1 ... \rho_{2p+1}}}
$$
={1\over{4\pi}} \epsilon^{\mu_1 ... \mu_{2p+1} \nu \rho_1 ...\rho_{2p+1}}
\int d^{4p+3} x A_{\mu_1 ... \mu_{2p+1}} 
\p_\nu A_{\rho_1 ... \rho_{2p+1}}$$ 
where the $(4p+3)$-dimensional integration is over the region
$x_{4p+2} \geq 0$. 

As discussed in \wit, there is a problem with obtaining the partition
function for a chiral $2p$-form from a path integral formalism. 
The problem is that the partition function depends on spin structure,
but there is no such dependence, at least naively, in the action of
\simp. This problem is related to the fact that, when the chiral $2p$-form
is coupled to a background $2p+1$-form
gauge field $A_{\mu_1 ... \mu_{2p+1}}$,
the partition function is not invariant under
gauge transformations of $A_{\mu_1 ... \mu_{2p+1}}$. 
To define the partition function in \wit, it was useful to
introduce a level $\half$
Chern-Simons action for $A_{\mu_1 ... \mu_{2p+1}}$, which
was precisely the action in \csg. This suggests that the Chern-Simons
form of the MWYW action might be useful for resolving the problem.

\newsec{Manifestly Covariant Actions for D=4 Self-Dual Yang-Mills}

\subsec{MWYW Action for Self-Dual Yang-Mills}

Self-dual Yang-Mills is described by a gauge field $A_\mu^I$ whose
field-strength $F_{\mu\nu}^I = \p_{[\mu} A_{\nu]}^I + f^I_{JK} A_\mu^J A_\nu^K$
is self-dual, i.e. 
\eqn\sd{F_{\mu\nu}^I = \half \e_{\mu\nu\rho\sigma} F^{\rho\sigma\, I}} 
where the spacetime signature is $(2,2)$. 
Although \sd\ can be obtained as an equation of motion from the 
actions of  \ref\don{S. Donaldson, Proc. Lond. Math. Soc. 50 (1985) 1\semi
V.P. Nair and J. Schiff, Phys. Lett. 246B (1990) 423.}
or  \ref\lez{A.N. Leznov, Theor. Math. Phys. 73 (1988) 1233\semi
A.N. Leznov and M.A. Mukhtarov, J. Math. Phys. 28 (1987) 2574\semi
A. Parkes, Phys. Lett. 286B (1992) 265.},
these actions are not SO(2,2) Lorentz-invariant and require the
four-dimensional spacetime to be Kahler. 

By replacing the chiral boson constraint with \sd, it is 
straightforward to use the MWYW method to construct a manifestly 
Lorentz-invariant action for self-dual Yang-Mills. The analogue of
\simp\ is 
\eqn\simpy{{\cal S}=\int d^4 x Tr [{1\over 2} F^{\mu\nu }F_{\mu\nu} +
G^{\mu\nu}_{(0)} (F_{\mu\nu} -{1\over 2}\e_{\mu\nu\rho\sigma} F^{\rho\sigma}) 
+\sum_{n=0}^\infty (-1)^n G^{\mu\nu}_{(n)} G_{(n+1)\mu\nu}]}
where $G_{(n)\mu\nu}^I $
are  Lagrange multipliers which will be taken to be
anti-self-dual, i.e. 
$G_{(n)\mu\nu}^I = -\half \e_{\mu\nu\rho\sigma} G_{(n)}^{\rho\sigma\, I}.$
Unlike the actions of \don\ and
\lez, \simpy\ can be generalized to any four-dimensional background. 
Note that the first terms in the action are the same as the action
of  \ref\si{G. Chalmers and W. Siegel, Phys. Rev. D54 (1996) 7628.},
and the infinite sum removes the propagating field in $G_{(0)}^{\mu\nu\,I}$.
 
The equations of motion for \simpy\ are easily found to be
\eqn\soly{
(\p^\mu \d_J^I +
f^I_{JK} A^{\mu\,K})
[F_{\mu\nu}^J+ 2 G_{{(0)}\mu\nu}^J  ]=0,\quad 
F_{\mu\nu}^I =\half\e_{\mu\nu\rho\sigma} F^{\rho\sigma\,I} -
G^{\mu\nu\,I}_{(1)} ,} 
$$G^{\mu\nu\,I}_{(0)} = 
G^{\mu\nu\,I}_{(2)} = ..., \quad 
G^{\mu\nu\,I}_{(1)} = 
G^{\mu\nu\,I}_{(3)} = .... $$ 
If the solutions to \soly\ 
are required to contain a finite number of non-vanishing
fields, the solutions must satisfy
\eqn\saty{
F_{\mu\nu}^I =\half\e_{\mu\nu\rho\sigma} F^{\rho\sigma\,I} 
,\quad G^{\mu\nu\,I}_{(n)} =0}
which are the desired conditions.   

\subsec {Self-Dual Maxwell Theory from Discrete Form of a 5D Action}

The analogue of
\simp\ for self-dual  Yang-Mills is 
 \simpy, and just as \simp\ can be obtained by gauge-fixing the action \cov, 
\simpy\ can be obtained by gauge-fixing an
action, at least in the abelian case. 
The action \simpy\
 can be written
in form notation as
\eqn\simpym{{\cal S}=\int   [{1\over 2} 
F \we *F + 
{1\over 2}  
G _{(0)}\we  (*F-F)
+\sum_{n=0}^\infty (-1)^n G _{(n)}\we  * G_{(n+1) }]}
In the abelian case, this can be obtained by gauge-fixing the following action:
\eqn\simpyn{{\cal S}=\int  \sum_{n=0}^\infty [{1\over 2} 
(-1)^n dB_{(n)} \we *dB_{(n)}
+ (G _{(2n)}-G _{(2n+2)})
\we * G_{(2n+1) } }
$$ +G _{(2n)}\we *d(B _{(2n)}+B _{(2n+1)})
-G _{(2n+1)}\we *d(B _{(2n+1)}+B _{(2n+2)})
]$$
where the $B _{(n)}$ are 1-forms, with
$B _{(0)}=A$, $dB _{(0)}=F$.
\simpyn\ is invariant under the symmetries:
\eqn\gauges{\delta 
B_{(n)}=\l_{(n-1)}-\l_{(n)},
\quad\delta G _{(n)}= {1\over 2} [d \l_{(n)}-*d \l_{(n)}],} 
(with $\l_{(-1)} \equiv 0$) where the parameters $\l_{(n)}$ are now 1-forms.
This allows $B_{(n)}$ to be algebraically 
gauged away for $n\geq 1$, which produces the action \simpy, 
where ${1\over 2}  
G _{(0)}\we  (*F-F)=G _{(0)}\we  *F$ 
since $G_{(0)}$ is anti-self-dual.

The chiral boson action was seen to be related to a 3-dimensional 
Chern-Simons theory, and the formal similarity with that case suggests seeking
a 5-dimensional theory whose discretisation gives an action of this
type. A simple guess would be to
define five-dimensional fields 
$$C^-(x^\mu, 2n\d x^4)=G_{(2n)} (x^\mu),\quad  
C^+ (x^\mu, (2n+1)\d x^4)=G_{(2n+1)}(x^\mu), $$
$$
C (x^\mu, n\d x^4) =B_{(n)}(x^\mu)/\delta x^4.$$
In  
the continuum limit $\delta x^4\to 0$, \simpyn\ becomes
\eqn\conti{{\cal S}= \epsilon^{\mu\nu\rho\sigma}
\int d^5 x [ C^+_{\mu\nu} \p_4 C^-_{\rho\sigma} 
+C^-_{\mu\nu} \p_\rho C_{\sigma} 
-C^+_{\mu\nu} \p_\rho C_{\sigma} ] .}
However, the anti-self-duality of 
$C^\pm_{\mu \nu}$ is not a covariant constraint in five dimensions,
so \conti\ is not SO(3,2) Lorentz-covariant. 

To get a Lorentz-covariant action, one could try writing
an alternative action involving infinite fields
in which the multiplier fields $G_{(n)}$
are not restricted to be
anti-self-dual. For example, the action 
\eqn\simpyt {{\cal S}=\int  \sum_{n=0}^\infty [{1\over 2} 
(-1)^n dB_{(n)} \we *dB_{(n)} 
+ (G _{(2n)}-G _{(2n+2)})
\we * G_{(2n+1) } }
$$ +G _{(2n)}\we *d(B _{(2n)}+B _{(2n+1)})
-G _{(2n+1)}\we *d(B _{(2n+1)}+B _{(2n+2)})
]$$
is invariant under
\eqn\gaugest{\delta 
B_{(n)}=\l_{(n-1)}-\l_{(n)},\quad\delta G _{(n)}=  d \l_{(n)} ,} 
for general 2-forms $G_{(n)}$. But on gauge-fixing, it gives
$G _{(0)}\we  *F$ instead of  $  
G _{(0)}\we  (*F-F)$.
One way of obtaining 
$G _{(0)}\we  (*F-F)$
would be to add to \simpyt\ the  term
\eqn\simpymod {{\cal S}'=\int  \sum_{n=0}^\infty {1\over 2} 
G_{(0)}\we  dB_{(n)}  }
(which is gauge-invariant up to a surface term), but \simpymod is 
non-local in the fifth dimension. Note that 
the simplest guess for a covariant five-dimensional 
limit of \simpyt\ would be $\int CdC$, but such an
action for a 2-form  $C$ is trivial since the integrand is a total derivative. 

\newsec{Manifestly Supersymmetric Actions for D=10 super-Yang-Mills } 

\subsec{MWYW Action for D=10 super-Yang-Mills } 

The physical fields of ten-dimensional super-Yang-Mills consist
of a gauge field $A_\mu^I$ and a Majorana-Weyl spinor field
$\Psi_\alpha^I$ where $I$ takes values in the adjoint representation.
On-shell, these fields satisfy the equations of motion 
\eqn\ons{{\cal D}_\mu F^{\mu\nu} = 0,\quad \Gamma_\mu^{\a\b} 
{\cal D}^\mu \Psi_\a =0,}
which can be obtained from the action
\eqn\sym{\int d^{10}x Tr ( \half F^{\mu\nu} F_{\mu\nu}
+i\Psi_\a \Gamma_\mu^{\a\b} {\cal D}^\mu \Psi_\b )}
where ${\cal D}^\mu \Psi_\b = \p^\mu \Psi_\b + [A^\mu, \Psi_\b]$ 
and $\Gamma_\mu^{\a\b}$ are the symmetric $16\times 16 $ Pauli matrices
in ten dimensions. 
Although this action is invariant under the global supersymmetry
transformations
\eqn\susy{\d A_\mu^I = i\e_\a \Gamma_\mu^{\a\b} \Psi_\b, \quad
\d \Psi_\a = -\half F^{\mu\nu} (\Gamma_{\mu\nu})_\a^\b \e_\b,}
this invariance is not manifest. 

It is well-known  \ref\nils{B.E.W. Nilsson, ``Off-shell Fields for
the Ten-Dimensional Supersymmetric Yang-Mills Theory'', 
Gothenburg preprint 81-6 (Feb. 1981), unpublished\semi
B.E.W. Nilsson, Class. Quant. Grav. 3 (1986) L41\semi 
E. Witten, Nucl. Phys. B266 (1986) 245.} 
that the equations of motion of \ons\ 
can be written in manifestly supersymmetric notation as
\eqn\sup{D^\a \Gamma^{\mu_1 ... \mu_5}_{\a\b} {\cal A}^{\b} (x,\t) = 0}
where $\t^\a$ is a Majorana-Weyl spinor variable, $\Gamma^{\mu_1 ... \mu_5}$
is the anti-symmetrized product of five $\Gamma$-matrices,  
${\cal A}^\a$ is a spinor superfield whose component expansion is
\eqn\comp{{\cal A}^\a = \xi^\a +  \Gamma_\mu^{\a\b}\t_\b  A^\mu
+  \Gamma^{\mu\,\a\b}\t_\b (\t_\g \Gamma_\mu^{\g\d} \Psi_\d)  + ... ,}
and 
$$D^\a \Gamma^{\mu_1 ... \mu_5}_{\a\b} {\cal A}^\b = 
\Gamma^{\mu_1 ...\mu_5}_{\a\b}
\left[ \left( 
{\p\over{\p\t_\a}} + i\t_\g \Gamma_\mu^{\a\g} \right) {\cal A}^\b +
\{ {\cal A}^\a, {\cal A}^\b \} \right].$$ 

Note that \sup\ is invariant under $\d{\cal A}^\a = D^\a \Lambda$ for
an arbitrary superfield $\Lambda$, which allows the component field
$\xi^\a$ to be gauged away. In this gauge, 
\sup\ implies that the higher components 
in the $\theta$ expansion of ${\cal A}_\a$ 
are all related to $\psi_\b$ and $A^\mu$.  

The natural generalization of \simp\ and \simpy\ for this case is
the action
\eqn\simps{
{\cal S}=\int d^{10} x \int d^{16}\t  ~Tr [
G_{(0)\,\mu_1 ... \mu_5}  
D^\a \Gamma^{\mu_1 ... \mu_5}_{\a\b} {\cal A}^{\b} 
+\sum_{n=0}^\infty (-1)^n
G^{\mu_1 ...\mu_5}_{(n)} G_{(n+1)\mu_1 ... \mu_5}]} 
where $G^{\mu_1 ... \mu_5}_{(n)}$ is an unconstrained superfield which
takes values in the adjoint representation. Note that one could also
include a quadratic term for ${\cal A}^\alpha$ of the type 
$ (D^\a \Gamma^{\mu_1 ... \mu_5}_{\a\b} {\cal A}^{\b})
(D^\g \Gamma_{{\mu_1 ... \mu_5}\,{\g\d}} {\cal A}^{\d})$, but such
a term vanishes in ten dimensions.\foot{We would like to thank 
Mario Tonin for pointing this out to us.} 
   
The equations of motion from varying the superfields in \simps\ are 
\eqn\sols{
D^\a \Gamma^{\mu_1 ... \mu_5}_{\a\b}
G_{{(0)}\mu_1 ... \mu_5} =0, \quad 
D^\a \Gamma^{\mu_1 ... \mu_5}_{\a\b} {\cal A}^{\b} +
G_{(1)}^{\mu_1 ... \mu_5} 
=0, } 
$$G^{\mu_1 ... \mu_5}_{(0)} = 
G^{\mu_1 ... \mu_5}_{(2)} = ..., \quad 
G^{\mu_1 ... \mu_5}_{(1)} = 
G^{\mu_1 ... \mu_5}_{(3)} = ...  .$$
If solutions to \sols\  
are required to contain a finite number of non-vanishing
fields, the solutions must satisfy
\eqn\saty{
D^\a \Gamma^{\mu_1 ... \mu_5}_{\a\b} {\cal A}^{\b} =0,  
\quad 
G^{\mu_1 ... \mu_5}_{(n)} = 0,}  
which are the desired conditions.   

\subsec{Super-Maxwell Theory from Discrete Form of an 11D Action} 

Just like the actions of \simp\ and \simpy\ , 
the action of \simps\ in the abelian case 
can be obtained by
gauge-fixing the following action:  
\eqn\ssc{
{\cal S}=\int d^{10} x \int d^{16}\t  \sum_{n=0}^\infty  [ 
(G_{(2n)} -G_{(2n+2)}) \wedge
*G_{(2n+1)} }
$$+ 
G_{(2n)}\wedge  ( *H_{(2n)} + *H_{(2n+1)} ) 
- G_{(2n+1)}\wedge ( *H_{(2n+1)} + *H_{(2n+2)} ) ] $$ 
where 
$H_{(n)}^{\mu_1 ... \mu_5} = 
D^\a \Gamma^{\mu_1 ... \mu_5}_{\a\b} {\cal A}_{(n)}^{\b}$, 
${\cal A}_{(0)}^\b$ is the original gauge superfield, and we
have switched to form notation. (Note that $H_{(n)}\wedge *H_{(n)}$
vanishes identically in ten dimensions.) 

This action is invariant under the gauge transformations
\eqn\gaugess{\delta 
{\cal A}^\a_{(n)}=\l^\a_{(n-1)}-\l^\a_{(n)},\quad
\delta G^{\mu_1 ... \mu_5}_{(n)}=   
D^\a \Gamma^{\mu_1 ... \mu_5}_{\a\b} {\l}_{(n)}^{\b} ,}
which can be used to algebraically gauge-fix ${\cal A}^\a_{(n)} =0$ 
for $n\geq 1$, which returns \ssc\ to the action of \simps. 

Defining  
$$C^- (x^\nu, 2n\d x^{10})=G_{(2n)},\quad  
C^+ (x^\nu, (2n+1)\d x^{10})=G_{(2n+1)} (x^\nu ),$$ 
$$
{\cal C}^\a (x^\nu, n\d x^{10})= {\cal A}^\a_{(n)}(x^\nu)/\delta x^4,$$
one can write \ssc\ in the limit $
\d x^{10} \to 0$ as the eleven-dimensional action 
\eqn\conts{{\cal S}= 
\int d^{11} x [ C^+_{\mu_1 ... \mu_5} \p_{10} C^{-\, \mu_1 ... \mu_5} 
+(C^-_{\mu_1 ... \mu_5} - C^+_{\mu_1 ... \mu_5} )
D^\a \Gamma^{\mu_1 ... \mu_5}_{\a\b} {\cal C}^\b]  .}
Of course, this action is not SO(10,1) invariant since $\mu$ ranges
from 0 to 9 and $\a$ ranges from 1 to 16. 

\newsec{Comments on Infinite Fields}

In this paper, we constructed manifestly covariant actions for
D=4 self-dual Yang-Mills and D=10 super-Yang-Mills theories using an infinite
number of fields. Although actions involving an infinite number of
fields are not often used, there are several instances where they
naturally arise. 

One instance was already mentioned and involves
discretizing a spacetime dimension. It would be interesting
to try to generalize the actions of \cov, \simpyn\ and
\ssc\ to the non-abelian case. 

A second instance of actions involving infinite fields 
is that of a Kaluza-Klein reduction, in which 
a Fourier expansion in the coordinate
of a compact dimension leads to an infinite number of fields.

A third instance of actions involving infinite fields are the
harmonic superspace actions of  \ref\harm{A. Galperin, E. Ivanov,
S. Kalitzin, V. Ogievetsky and E. Sokatchev, Class. Quant. Grav. 1 (1984)
469.}. The fields in these actions depend polynomially on a bosonic
``harmonic'' variable $u$, and Taylor expanding in $u$ produces
the infinite fields. 
In the superstring field theory action for Ramond-Ramond fields, which
is of the MWYW type, the infinite fields arise in precisely this manner
where the harmonic variable
$u$ is constructed from a combination of the bosonic ghost zero modes
\sft.
Perhaps the infinite fields in the actions of \simpy\ and \simp\ can
also be interpreted as a Taylor expansion in some harmonic variable.
This would connect with earlier attempts to construct actions for
D=4 self-dual Yang-Mills and D=10 super-Yang-Mills using harmonic variables
 \ref\sdy{S. Kalitzin and E. Sokatchev, Phys. Lett. B257 (1991) 151\semi
N. Marcus, Y. Oz and S. Yankielowicz, Nucl. Phys. B379 (1992) 121.} 
 \ref\sds{E. Sokatchev, Phys. Lett. B169 (1986) 209\semi
P. Howe, Phys. Lett. B258 (1991) 141, add.-ibid. B259 (1991) 511. }.

\vskip 30pt

{\bf Acknowledgements:} 
NB would like to 
thank N. Nekrassov for discussions on discretized Chern-Simons,
and Queen Mary and Westfield College for their hospitality. 
The
work of NB was partially supported by 
CNPq grant number 300256/94.

\listrefs
\end